\documentclass[12pt,preprint,aps,floatfix,onecolumn]{revtex4}
\usepackage{epsfig,amsmath,bbm,graphicx,color,pdfcolmk}
\usepackage[normalem]{ulem}

\begin{document}

\title{ A structure-based model fails to probe  the mechanical unfolding pathways 
of  the titin I27 domain}

\author{Maksim Kouza}
\email{mkouza@chem.uw.edu.pl} 
\affiliation{Faculty of Chemistry, University of Warsaw, Pasteura 1 02-093 Warsaw, Poland}

\author{Chin-Kun Hu}
\email{huck@phys.sinica.edu.tw} 
\affiliation{Institute of Physics, Academia Sinica, Nankang, Taipei 11529, Taiwan}

\author{Mai Suan Li}
\email{masli@ifpan.edu.pl} 
\affiliation{Institute of Physics, Polish Academy of Science, Al. Lotnikow 32/46 02-668
Warsaw, Poland}

\author{Andrzej Kolinski}
\email{kolinski@chem.uw.edu.pl}
\noaffiliation
\affiliation{Faculty of Chemistry, University of Warsaw, Pasteura 1 02-093 Warsaw, Poland}
\noaffiliation

\date{\today}

\begin{abstract}
We discuss the use of  a structure based C$\alpha$-Go model and Langevin
dynamics to study in detail  the mechanical properties and unfolding pathway of  the 
titin I27 domain. We show that  a simple Go-model does detect correctly the
origin of the mechanical stability of this domain. The unfolding free energy
landscape parameters $x_u$ and $\Delta G^{\ddagger}$, extracted from
dependencies of unfolding forces on pulling speeds, are found to agree
reasonably well with  experiments. We predict that above $v=10^4$ nm/s the
additional force-induced intermediate state is populated at  an end-to-end
extension of about $75 \mathring{A}$. The force-induced switch in  the unfolding
pathway occurs at the critical pulling speed $v_{crit} \approx 10^6-10^7$
nm/s. We argue that this critical pulling speed is an upper limit of the
interval where Bell's theory works. However, our results suggest that
 the Go-model fails to reproduce the experimentally observed mechanical unfolding
pathway properly, yielding an incomplete picture of the free energy
landscape. Surprisingly, the experimentally observed intermediate state with  the
A strand  detached is not populated in Go-model simulations over a wide
range of pulling speeds. The discrepancy between simulation and experiment
is clearly seen from the early stage of  the unfolding process  which shows the
limitation of the Go model in reproducing unfolding  pathways and
deciphering the complete picture of the free energy landscape.
\end{abstract}

\maketitle

\section{Introduction}

\begin{figure}
\includegraphics[angle=0,scale=1]{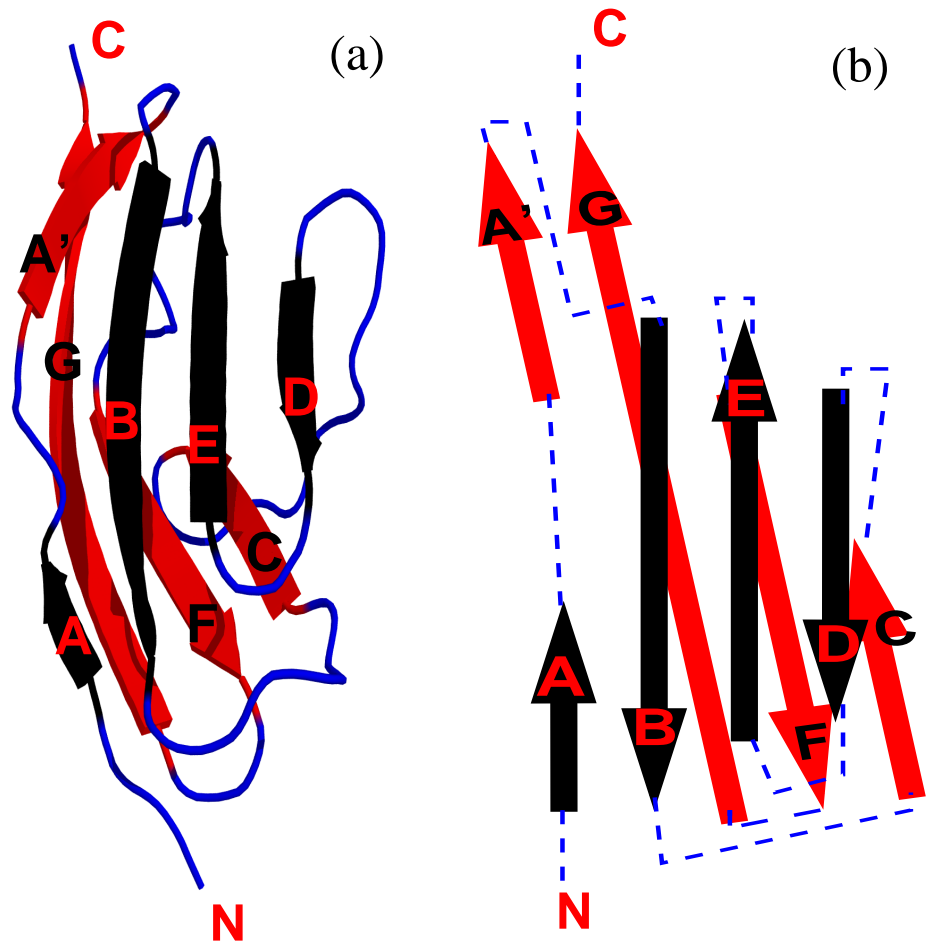}
\caption{ Native state conformation of  the I27 domain of titin (PDB ID: 1TIT).
		There are 8 $\protect\beta$-strands: A (4-7), A' (11-15), B (18-25), C
		(32-36), D (47-52), E (55-61), F(69-75) and G (78-88).  (a) PDB structure in
		 cartoon representation. (b) Schematic view of  the same structure. Red and black $%
		\protect\beta$-strands belong to different $\protect\beta$-sheets. N- and C-terminal residues are marked N and C, respectively.}
\label{Fig1_NS}
\end{figure}

Over the last 15 years the mechanical unfolding of  the titin  I27 domain has been
a subject of intense experimental and theoretical studies \cite%
{Tskhovrebova_Nature97, Rief_Science97,Lu_BJ98, Marszalek_Nature99,
Fowler_JMB02, Lee_Structure09}. 
This domain consists of 8 $\beta $-strands (Fig. \ref{Fig1_NS}) that fold
into two layers of $\beta $-sheets through backbone hydrogen bonds (HBs) and
side-chain interactions and it has high resistance to external force. One of
the most remarkable findings is that the force-extension profile of I27
displays a "hump",  a previously overlooked
pre-peak preceding the main peak\cite{Marszalek_Nature99}. It was interpreted by steered
molecular dynamics (SMD) all-atom simulation in explicit solvent as a
signature of  an unfolding intermediate in which all hydrogen bonds between  the A
and B strands  were  broken\cite{Lu_CP99, Lu_BJ00}. Experimentally, this
suggestion was confirmed by studies where a Lys to Pro point mutation at
residue 6 
disrupted hydrogen bonds connecting strand A with B and resulting peaks on
the force-extension profile occur red without humps\cite{Marszalek_Nature99}.
Two mutants,  the I27 domain with  a detached A strand and  a destabilized A strand by  a 
Val to Ala mutation at residue 4 have been used to verify whether the
unfolding behaviour would be affected. Both of them did not show the
difference between  the wild type  and mutants of  the I27 domain in unfolding forces  or  in the
dependencies of unfolding forces on pulling speed\cite{Fowler_JMB02},
suggesting the mechanical unfolding of titin as a two-step process\cite%
{Marszalek_Nature99,Fowler_JMB02}.
The first step is  a transition from  the native (N) to intermediate state (IS),
which corresponds to breaking of HBs between $\beta $-strands A and B and
unfolding of A at the end-to-end extension $\approx 7\mathring{A}$.  A force  of about $100$ pN is necessary to cross the first transition state (TS1) and
form a stable intermediate structure with  the A strand detached \cite%
{Marszalek_Nature99}. The second stage,  a transition from IS to the denatured
state (DS), was initially believed to be solely associated with the
cooperative rupture of six HBs between strands A' and G. 
 Subsequently , it  was shown that together with hydrogen bonding interactions
the side-chain packing in  the A'G region plays an important role in  the unfolding
process and contributes to  the mechanical stability of titin \cite{Fowler_JMB02,
Best_JMB03}.  A force  of about  $200$ pN is required to cross the second
transition state (TS2) and unfold the protein  completely. Despite the
important role of water molecules\cite{Lu_BJ00}, implicit solvent
simulations also revealed  the detachment of  the A strand as  the first step in
 the unfolding process \cite{Paci_PNAS00, Fowler_JMB02}. Thus,  until now
detailed all-atom simulations using  the CHARMM force field either with  the TIP3P
water model\cite{Lu_BJ98,Lu_CP99} or a continuum representation of solvent%
\cite{Fowler_JMB02, Paci_PNAS00} as well as experiments\cite%
{Marszalek_Nature99} show ed that structure A unfolds first. However,
deciphering the unfolding free energy landscape (FEL) of long proteins by
all-atom simulations with explicit solvent is still
computationally inaccessible. The time scale discrepancy (and the
discrepancy in stretching forces required to induce unfolding) between AFM
experiments and simulations can be reduced  using Go models\cite%
{Go_ARBB83, Clementi_JMB00}.
They have been successful in
describing  the folding and unfolding of a number of proteins
\cite{Hills_IJMS09, Estacio_JCP12, 
Kouza_JCP11,
West_JCP06, 
Kouza_JCP08, Cieplak_Proteins02, Sharma_Bioinformatics09, Sulkowska_Proteins08,Schlierf_BJ10, Yew_JPCB08},
but mechanical unfolding studies  typically  agree with experimental data  in
mechanostability properties, such as the unfolding force ($F_{max}$) and FEL
parameters including the distance between the native and transition states $%
x_{u}$ and the unfolding barrier $\Delta G^{\ddagger }$. Little success has
been obtained for mechanical unfolding pathways as pathways of titin,
Ubiquitin and DDFLN4 predicted by Go-model\cite{Cieplak_Proteins02,
MSLi_BJ07, Li_JCP09} did not agree with experimental findings. In the case
of Ubiquitin,  the intermediate state  was overlooked due to the lack of
non-native interactions.  The incorporation of these interactions helped to
detect it correctly \cite{Irback_PNAS05}.  An analogous result was obtained
for protein DDFLN4,  in which  non-native interactions  were shown to lead to
an intermediate state\cite{Kouza_JCP09} previously undetected by Go-model
simulations\cite{Li_JCP09}.

 For titin, it is unclear why  the intermediate state can not be
captured by Go-model, as 
an inaccurate pathway  was obtained not only by  Go-model\cite%
{Cieplak_Proteins02} but also  by an all-atom simulation in implicit  solvent%
\cite{Li_JCP03},  in which non-native interactions  were taken into account.
Complete breaking of A'G contacts  was shown to occur before the rupture
of contacts between strands A and B. Consequently, it fail ed to capture  the
experimentally observed intermediate state 
as the crossing over the first transition state should be associated with a 
loss of native interactions or breaking of HBs between  the A and B strands. On
the other hand, it has been recently shown that extreme conditions change
the unfolding pathway of other $\beta$-strand proteins (FnIII and DDFLN4) \cite%
{Mitternacht_BJ09, Li_JCP09}. Therefore, one of the possible reasons  for
the difference in titin unfolding pathways  is that pulling speed, $%
v=17\times10^6$ nm/ps, used in  the simulations \cite{Cieplak_Proteins02} is  a few
orders of magnitude higher than that  used in AFM experiments \cite%
{Carrion-Vasquez_PNAS99}.

 An interesting question arises whether  the Go-model could correctly describe
the unfolding pathways of the best-studied titin I27 domain at pulling
speeds close to the experimental ones. In the present paper we address this
question using the Go model version developed in Ref. \cite%
{Clementi_JMB00}. We found that the second peak that occurs at  an 
end-to-end extension  of $\approx 70$ $\mathring{A}$ disappears at pulling
speeds  lower than $v=10^{4}$ nm/s. The force-induced switch in  the unfolding
pathway  was shown to occur at the critical pulling speed $%
v_{crit}\approx 10^{6}-10^{7}$ nm/s. We propose that this critical pulling
speed constitutes an upper limit of the interval  in which  Bell's
theory works. It is shown that unfolding pathways depend on pulling speeds
but the Go model fails to describe the pathway observed in the
experiments even at pulling speeds comparable  to those used in
experiments. To summarize, contrary to the common belief that
structure-based models reproduce properly the key features of unfolding
process, the results obtained by Go-model simulations should be taken with a
grain of salt.

\section{Materials and methods}

\subsection*{Off-lattice Go model and Langevin dynamics}

We use coarse-grained continuum representation for  the I27 domain in which only
the positions of C$_{\alpha}$-carbons are retained. The interactions between
residues are assumed to be Go-like and the energy of such a model is as
follows \cite{Clementi_JMB00}

\begin{eqnarray}
E \; = \; \sum_{bonds} K_r (r_i - r_{0i})^2 + \sum_{angles} K_{\theta}
(\theta_i - \theta_{0i})^2 + \sum_{dihedral} \{ K_{\phi}^{(1)} [1 - \cos
(\phi_i - \phi_{0i})] +  \notag \\
K_{\phi}^{(3)} [1 - \cos 3(\phi_i - \phi_{0i})] \} + \sum_{i>j-3}^{NC}
\epsilon_H \left[ 5\left( \frac{r_{0ij}}{r_{ij}} \right)^{12} - 6 \left( 
\frac{r_{0ij}}{r_{ij}}\right)^{10}\right] + \sum_{i>j-3}^{NNC} \epsilon_H
\left(\frac{C}{r_{ij}}\right)^{12} .  \label{Hamiltonian}
\end{eqnarray}
Here $\Delta \phi_i=\phi_i - \phi_{0i}$, $R_{ij}={r_{0ij}}/{r_{ij}}$; $%
r_{i,i+1}$ is the distance between beads $i$ and $i+1$, $\theta_i$ is the
bond angle  between bonds $(i-1)$ and $i$, and $\phi_i$ is the dihedral
angle around the $i$th bond and $r_{ij}$ is the distance between the $i$th
and $j$th residues. Subscripts ``0'', ``NC'' and ``NNC'' refer to the native
conformation, native contacts and non-native contacts, respectively.
Residues $i$ and $j$ are in native contact if $r_{0ij}$ is less than a
cutoff distance $d_c$ taken to be $d_c = 6$ \AA , where $r_{0ij}$ is the
distance between the residues in the native conformation. With this choice
of $d_c$ the native conformation from the PDB we have 86 native contacts in
total.

The first harmonic term in Eq. (\ref{Hamiltonian}) accounts for chain
connectivity and the second term represents the bond angle potential. The
potential for the dihedral angle degrees of freedom is given by the third
term in Eq. (\ref{Hamiltonian}). The interaction energy between residues
that are separated by at least 3 beads is given by 10-12 Lennard-Jones
potential. A soft sphere (last term in Eq. (\ref{Hamiltonian})) repulsive
potential disfavors the formation of non-native contacts. We choose $K_r =
100 \epsilon _H/\mathring{A}^2, K_{\theta} = 20 \epsilon _H/rad^2, 
K_{\phi}^{(1)} = \epsilon _H$, and $K_{\phi}^{(3)} = 0.5 \epsilon _H$, where 
$\epsilon_H$ is the characteristic hydrogen bond energy and $C = 4$ \AA .
Since $T_F=0.5 \epsilon_H$ (see below) and $T_F = 333 K$ \cite%
{Politou_Biochemistry94}, we have $\epsilon_H=1.37$ kcal/mol. Then the force
unit $\epsilon_H/\mathring{A}\, = 95$ pN. The dynamics of the system is
obtained by integrating the following Langevin equation \cite%
{Allen_book,Kouza_BJ05} 
\begin{equation}
m\frac{d^2\vec{r}}{dt^2} \; \; = \; \; - \zeta \frac{d\vec{r}}{dt} + \vec{F}%
_c + \vec{\Gamma},  \label{DynaEq_eq}
\end{equation}
where $m$ is the mass of a bead, $\zeta$ is the friction coefficient, $\vec{F%
}_c = -dE/d\vec{r}$. The random force $\vec{\Gamma}$ is a white noise, i.e. $%
<\Gamma (t) \Gamma (t^{\prime})> = 2\zeta k_BT\delta(t-t^{\prime})$. It
should be noted that the folding thermodynamics does not depend on the
environment viscosity (or on $\zeta$) but the folding kinetics depends on
it. Most of our simulations (if not stated otherwise) were performed at the
friction $\zeta = 2\frac{m}{\tau_L}$, where the folding is fast. Here $%
\tau_L = (ma^2/\epsilon_H)^{1/2} \approx 3$ ps. The  equations of motion
were integrated using the velocity form of the Verlet algorithm \cite%
{Veitshans_FD97} with the time step $\Delta t = 0.005 \tau_L$. In order to
check robustness of our predictions for unfolding pathways limited
computations were carried out for the friction $\zeta = 50\frac{m}{\tau_L}$
which is believed to correspond to the viscosity of water \cite%
{Veitshans_FD97}). In this overdamped limit we use the Euler method for
integration and the time step $\Delta t = 0.1 \tau_L$. Three types of
Langevin dynamics simulations were carried out. (i) In the absence of force.
(ii) A constant force was applied to both termini. In the later case one has
to add to the energy (\ref{Hamiltonian}) the term -$(\vec{f}\vec{r})$ where $%
\vec{r}$ is the end-to-end vector. (iii) In the constant velocity force
simulation we fix the N-terminal and pull the C-terminal by force, $%
f=K_r(vt-r)$, 
where r is the displacement of the pulled atom from its original position
and the spring constant of cantilever, $K_r$, is set to be the same as the
spring constant of the Go model. The pulling direction was chosen along the
vector connecting N- and C-terminal atoms.

\subsection*{Tools and measures used in the analysis.}
The temperature-force phase diagram and the thermodynamic quantities were
obtained by the multiple histogram method \cite{Ferrenberg_PRL89} extended
to the case when the external force is applied to the termini \cite%
{Klimov_PNAS00}. The reweighting is carried not only for temperature but
also for force. We collected data for five values of $T$ at $f=0$ and and
for five values of $f$ at a fixed value of $T$. The duration of MD runs for
collection of histograms was chosen to be the same for all trajectories. In
order to obtain sufficient sampling 30 independent trajectories were
generated at each value of temperature and force.

We studied the unfolding pathways by monitoring the fraction of native
contacts of each $\beta $-strand and their pairs as a function of end-to-end
distance which is believed to be a good reaction coordinate. All-atom
representations were obtained by reconstruction of the backbone and
side chain atoms by BBQ method\cite{Gront_JCC07} and SCWRL 4.0 package\cite%
{Krivov_Proteins09} described in more detail in Ref.\cite{Kmiecik_JPCB12}.

\section{Results}

\subsection*{Temperature-force phase diagram}

\begin{figure}
\includegraphics[width=4.0in]{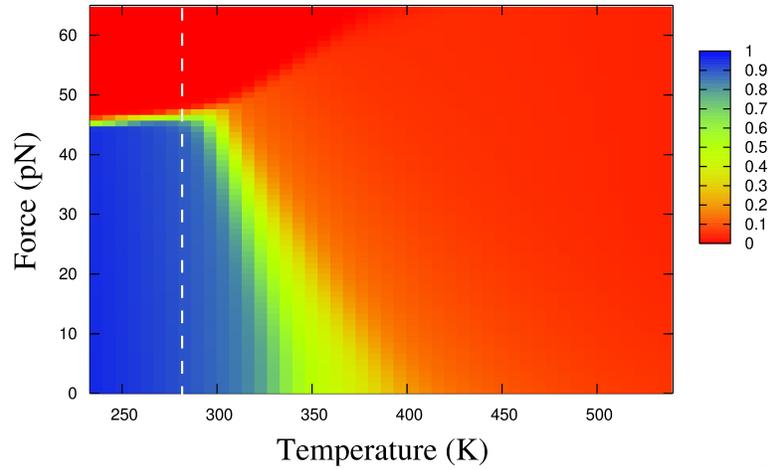} 
\caption{The $f-T$ phase diagram obtained by the extended histogram method.
		The results were averaged over 30 trajectories. The vertical dashed line
		marks $T=0.42\protect\epsilon _{H}/k_{B}=280$ K at which  most of our
		calculations have been performed.}
\label{Fig2_ns_phase_fig}
\end{figure}

The $f-T$ phase diagram, obtained  from the extended histogram method (see 
\emph{Materials and Methods}) is shown in Fig. \ref{Fig2_ns_phase_fig}. The
folding-unfolding transition is defined by the yellow region which is sharp
in the low temperature region but becomes less cooperative (the
fuzzy transition region is wider) as $T$ increases. The weak reentrancy
occurs at low temperatures, where a slight decrease  in the critical force
with $T$ is observed. This seemingly strange phenomenon occurs as a result
of competition between the energy gain and the entropy loss on stretching.
 A similar cold unzipping transition was also observed in a number of
models for heteropolymers \cite{Shakhnovich_PRE02} and for proteins \cite%
{MSLi_BJ07}. In the absence of force the folding temperature, $T_{F}$, at
which $df_{N}/dT$ is  maximum (results not shown), is equal to $%
T_{F}=0.5\epsilon _{H}/k_{B}$. Equating this value  with the experimental value 
$T_{F}=333$ K \cite{Politou_Biochemistry94} we can extract the energy scale $%
\epsilon _{H}$ which is given in \emph{Materials and Methods}. At $%
T=0.42\epsilon _{H}/k_{B}$ = 280 K,  in which our simulations  were carried
out, the equilibrium critical force $f_{eq}=0.42\epsilon _{H}/\mathring{A}%
\,\approx 40$ pN. This value is higher than the experimental estimate $%
f_{eq}\approx 18$ pN \cite{Rief_Science97}. 
Given the simplicity of the Go model we use, this agreement with the
experiments is considered reasonable.

\subsection*{Force-extension profile:  the second peak disappears at low pulling
speeds}

\begin{figure}
\includegraphics[width=4.0in]{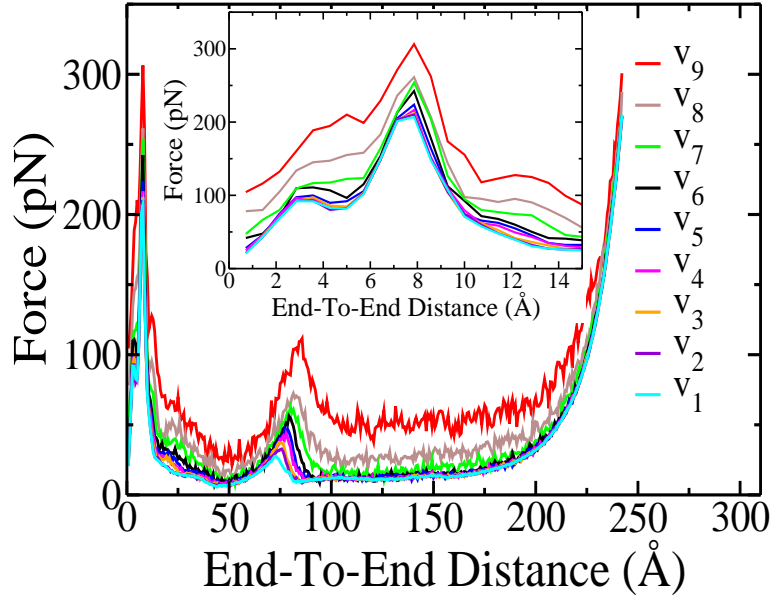} 
\caption{ Averaged force-extension profiles of  the I27 titin domain. The results obtained by Go-model simulations
		performed at different pulling speeds  are shown  on the right. The inset shows an
		enlargement of the starting  region. Results have been averaged over 50
		trajectories. Values of pulling rates are given in Table \protect\ref%
		{sim_details1}.}
\label{fe_profile_Go}
\end{figure}

\begin{figure}
\includegraphics[width=6.3in]{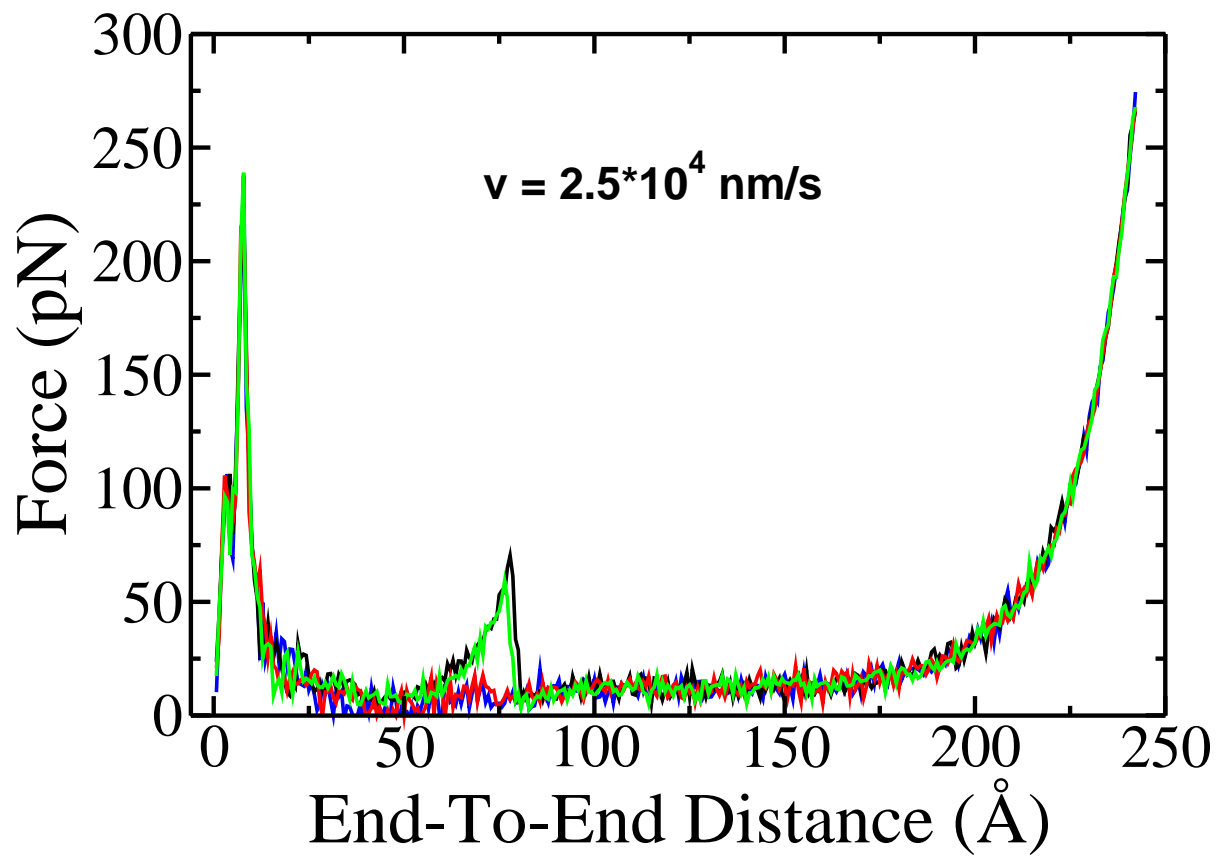} 
\caption{ Individual force-extension profiles of  the I27 titin domain. Four individual trajectories at $%
		2.5\times10^4$ nm/s obtained by Go-model simulations are presented. There is no second
		peak in two trajectories.}
\label{fe_profile_Go_individual}
\end{figure}

The force-extension profiles of I27 shown  in Fig. \ref{fe_profile_Go}
 have two peaks over a wide range of pulling speeds. The main peak occurs
at $\Delta R \approx 8 \mathring{A}$ for all  speeds studied, while the
position of the second lower peak depends on loading rates.  
As follows from Fig.\ref{fe_profile_Go} and Fig.\ref{fe_profile_Go_individual} the first peak is 
preceded by the hump 
which was also 
observed in AFM experiments
 and interpreted as a signature of  the intermediate state with  the A strand detached from the protein \mbox{\cite{Marszalek_Nature99}}. 
 At first glance, the existence of the hump and the first peak agrees well with  the AFM data,
 however, there is a substantial difference
in the nature of  the humps observed in  the experiments and  in our Go model. Since the hump
seen in  the experiments is caused by  the detachment of the whole A strand, the peak 
 associated with 
it may be considered as the first transition state (TS1) (Fig. S3 in
Supporting Materials). The hump observed in Go simulations, as shown below,
occurs due to breaking of only  two out of  the seven native contacts between
A and B and its smeared peak at  an extension of $4-5 \AA$ cannot be interpreted
as the transition state. Therefore, TS1 in the Go model is the main peak
at $\sim 8 \AA$ in
 Fig. \mbox{\ref{fe_profile_Go}} (see also Fig. S3 in SM), while the second peak
at $\sim 75 \AA$ corresponds to the second TS2. The experimental TS2
is the main peak located at $\sim 10 \AA$ (Fig. S3 in SM).

The second peak around
75-85 $\mathring{A}$  revealed by  the Go-model simulations indicates that an additional mechanical 
intermediate becomes populated. Thus,  the Go-model fails to predict  the experimentally observed 
intermediate state with  the A strand detached. Instead it predicts the existence of the peak around
75-85 $\mathring{A}$ far from the native conformation. This result was also
observed in  earlier Go-model  studies \mbox{\cite{Cieplak_Proteins02, MSLi_JCP08}} 
and confirmed by all-atom simulation in explicit solvent(see Fig. S1 in SM
\mbox{\cite{SuppInfo}}).
It should be noted that the height of  the second peak strongly decreases 
with  the decreasing pulling speed (Fig. \ref{fe_profile_Go}). Its average value
is as low as 36 pN at the lowest velocity $v_{1}=2.5\times 10^{4}$ nm/s.
More importantly, at this speed 30$\%$ of trajectories do not reveal
the second peak while at $v_{2}=5.76\times 10^{4}$ nm/s this  value becomes
18$\%$. For illustration we show four typical force-extension curves in Fig. %
\ref{fe_profile_Go_individual}, where trajectories shown in green and black
 proceed with clear second peaks, while those in red and green do not.
The increasing probability of such a pathway suggests that the second peak 
might vanish in  the experiment (see also below). As shown below, the Go
model does not correctly describe  the unfolding pathways even for trajectories
that proceed with the second peak.

\subsection*{Protein unfolding pathway dependence on pulling speed}

 To monitor unfolding sequencing we plot  the fraction of native contacts
formed by $\beta$-strands with the rest of  the protein and native contacts
formed by pairs of $\beta$-strands as a function of end-to-end distance, $%
\Delta R$. Fig.\ref{strands_pairs_Go} show s the $\Delta R$  dependence of
native contacts of all $\beta$-strands and their pairs for different pulling
speeds.

\emph{High pulling speed regime, $v\gtrsim 10^{6}$ nm/s}. In this regime,
where the pulling speed $v$ is larger than $10^{6}$ nm/s 
(Fig.\ref{strands_pairs_Go}a), unfolding starts from  the C-termin us (Eq.\ref%
{pathway_higha}). G and F strands are detached first followed by  the
simultaneous unfolding of strands A, A', B and C. Finally, unfolding of the
most stable strands E and D  occurs. Fig.\ref{strands_pairs_Go}d
gives the following sequence for interstrand contacts for  the high
velocity regime (Eq.\ref{pathway_highb}). Interstrand contacts begin to
break down from AG and AG' followed by  the serial breaking of FG, CF, BE and AB
contacts. Breaking  the DE contacts completes the unfolding process. The
typical unfolding pathway at $v=3.22\times 10^{6}$nm/s is shown  in  Fig. \ref%
{Go_model_path_all}(e-h). Clearly, that first peak  in the force-extension
profile at $\Delta R \approx 8 \AA$ corresponds to breaking  the AG and A'G contacts [Fig. \ref%
{strands_pairs_Go}(d-f)]. Once they are ruptured (Fig. \ref{Go_model_path}c),  the
protein passes from the transition state into the intermediate
one. A typical intermediate structure with  the C-terminus unfolded is shown  in Fig. \ref{Go_model_path_all}.

\begin{subequations}
\begin{equation}
G \rightarrow F \rightarrow (A, A^{\prime}, B, C) \rightarrow E \rightarrow
D,  \label{pathway_higha}
\end{equation}
\begin{equation}
AG \rightarrow A^{\prime}G \rightarrow FG \rightarrow CF \rightarrow BE
\rightarrow AB \rightarrow DE,  \label{pathway_highb}
\end{equation}

\emph{Low velocity regime, $v \lesssim 10^6$ nm/s}.  The unfolding  sequence below 
$v=2.88\times10^5$ nm/s  is different.  The G and F strands remain partially
structured before  the complete unfolding of  the A,A',B and C strands (Fig.\ref%
{strands_pairs_Go}c). Only after  the complete unfolding of  the A,A',B and C strands,  do 
F and G strands lose their secondary structure s (Eq.\ref{pathway_lowa}). We
observed  a switch in  the unfolding mechanism - below $v=2.88\times10^5$nm/s
 the unfolding starts form  the N-termin us, otherwise from the C-termin us. 
As follows from the dependence of intrastrand contacts  on $\Delta R$,  there is a
reverse order of events compared to  the high velocity  pulling regime, besides  the	
initial and final stages of the unfolding process. Namely,  the serial breaking
 AB, BE, CF, FG contacts proceeds after unraveling  the AG and A'G contacts,
but it precedes  the final unwrapping of  the DE contacts (Eq.\ref{pathway_lowb}).

\end{subequations}
\begin{subequations}
\begin{equation}
(A, A^{\prime}) \rightarrow B \rightarrow C \rightarrow G \rightarrow F
\rightarrow E \rightarrow D,  \label{pathway_lowa}
\end{equation}
\begin{equation}
AG \rightarrow A^{\prime}G \rightarrow AB \rightarrow BE \rightarrow CF
\rightarrow FG \rightarrow DE,  \label{pathway_lowb}
\end{equation}
\begin{equation}
(AG, AB) \rightarrow A^{\prime}G \rightarrow {?} ,  \label{pathway_experim}
\end{equation}

According to  the Go-model results, the first resistance point is the
contacts between  the A and A' with  the G  strand. 
Once it is ruptured , the force
drops drastically. The first peak  in the force-extension profile is robust for
all pulling speeds studied. The nature of second resistance point (peak) is
due to the core formed by either A,A',B,C strands at low pulling speed s or by
A,A',B,C, F strands at high pulling speeds. As contacts begin to break down
the further unraveling proceeds smoothly without significant resistance.

At first glance our Go-model data agree with the experiment where 
 at low loading rates strand A is detached from the protein first.
According to  the AFM experiment, AB HBs provide the first line of
defense against the external force. At about $100$ pN  the A strand
detaches out of  the G and B strands leading to  a $\sim 7-8\mathring{A}$ 
extension.  The second
line of defense lies in the A'G region. Once A'G HBs are broken the protein
no longer resists to the force. However, as seen in Fig. %
\ref{strands_pairs_Go}, the complete detachment of  the A strand takes place at  a $%
75\mathring{A}$ extension almost simultaneously with  the A' and B strands. Interestingly, similar ly to  the 
AFM experiment,  the Go-model shows 
the hump in the rising phase of the first peak (Fig.\ref{fe_profile_Go}). However, if we examine  the 
molecular origin of the hump, agreement between  the Go-model simulation and  the experiment is not observed. 
Namely, instead of  the full detachment of  the A strand (or breaking all HBs between  the 
A and B strands), we observe breaking 100$\%$  AG contacts  on average
(there are  two native contacts between A and G) and 
only up to 30$\%$ AB contacts (2 out of 7 native contacts between A and B). Thus,
unlike the experimentally confirmed unfolding pathway (Eq. \ref{pathway_experim}), we do not observe  the detachment of  the 
A strand at small
extensions within $10\mathring{A}$ over  the all pulling speeds  studied. If we
look at  the evolution of pairs (Fig. \ref{strands_pairs_Go}(d-f)), contacts
between  the A and B strands are always broken after those between A' and G.
Within  a $10\mathring{A}$ extension more than 70 $\%$ AB contacts remain
formed. Typical conformations at  an extension of 4 and 9 $\mathring{A}$ are shown
  in Fig. \ref{Go_model_path}b and \ref{Go_model_path}c, respectively. Thus the Go-model fails to reproduce the
experimental observation that  the A strand unfolds first regardless the applied
pulling force. In 100$\%$ trajectories the system is directed into an
alternative pathway, where the breaking of A'G and AG contacts takes place
first, while experiments showed that in the intermediate state  the A
strand should be detached from  the protein completely(or both AG and AB
contacts are broken). It is not clear whether the rupture events at  a larger
extension are correct,  as even at  a small extension the system is directed
into  the wrong pathway. Moreover, molecular interactions underlying the
mechanical resistance of  the protein might be altered. Fortunately, it is not
the case for titin, where  the mechanical resistance of the protein lies in  the A'G
region, while AB contacts do not contribute to mechanostability \cite{Fowler_JMB02, Best_JMB03}. However, one has to keep in mind, that, in
general,  the unfolding pathway probed by Go-models can differ from the pathway
studied in AFM experiments and  the molecular basis of mechanical stability might
be affected.

It is worth mentioning that all-atom simulations in explicit solvent, show correctly not only the  hump and but also 
the molecular structure behind it indicating the presence of  the experimentally observed intermediate state. All hydrogen bonds between 
 the A and B strands are broken  
after the protein passes the first transition state. 
Snapshots are shown in Fig. S2 in SM \mbox{\cite{SuppInfo}}.

\subsection*{FEL  parameters: Bell approximation and beyond}

In experiments one usually uses the Bell formula \cite{Bell_Science78} 
\end{subequations}
\begin{equation}
\tau_{U} = \tau_{U}^0\exp(-x_uf/k_BT)  \label{Bell_eq}
\end{equation}
to extract $x_u$ for two-state proteins from the force dependence  of
unfolding times $\tau_{U}$. Eq. \ref{Bell_eq} is valid if the location of
the transition state does not move under external force. Assuming that the
force increases linearly with pulling speed $v$ and the $x_u$ does not
depend on the external force,  Evans and Ritchie \cite{Evans_BJ97} have
shown that the distribution of unfolding force $P(f)$ obeys the following
equation: 
\begin{equation}
P(f) \; = \; \frac{k_{u}(f)}{v} \exp \{ \frac{k_BT}{x_uv} \left[
k_u(0)-k_u(f) \right]\},
\end{equation}
where $k_u(f) = \tau_U^{-1}$ is given by Eq. \ref{Bell_eq}. Then, the most
probable unbinding force or the maximum of force distribution  $f_{max}$,
obtained from the condition $dP(f)/df|_{f=f_{max}}=0$, is 
\begin{equation}
f_{max}=\frac{k_BT}{x_u}ln\frac{kvx_u}{k_u(0)k_BT}  \label{f_logV_eq}
\end{equation}

Schlierf and Rief have shown that if location of transition state is sensitive to the applied force, one has
to go beyond the Bell approximation \cite{Schlierf_BJ06}. Dudko \emph{et al} have proposed the
following force dependence for the unfolding time \cite{Dudko_PRL06}:

\begin{equation}
\tau _{u}\;=\;\tau _{u}^{0}\left( 1-\frac{\nu x_{u}}{\Delta G^{\ddagger }}%
\right) ^{1-1/\nu }\exp \{-\frac{\Delta G^{\ddagger }}{k_{B}T}[1-(1-\nu
x_{u}f/\Delta G^{\ddagger })^{1/\nu }]\}.
\label{Dudko_eq}
\end{equation}%
Here, $\Delta G^{\ddagger }$ is the unfolding barrier, and $\nu =1/2$ and
2/3 for the cusp \cite{Hummer_BJ03} and the linear-cubic free energy surface 
\cite{Dudko_PNAS03}, respectively. $\nu =1$ leads to the phenomenological
Bell theory (Eq. (\ref{Bell_eq})). Note that if $\nu \neq 1$,
both $x_{u}$ and $\Delta G$ can be determined.

Fig. \ref{unfold_barrier_CV} shows the most probable unbinding force as a
function of pulling speed. As evident from the plot, there exists  a
 critical  speed $v_{c}\approx 10^{6}-10^{7}$ nm/s , separating the low and
high pulling speed regimes. In the low pulling speed regime ($v\lesssim
4\ast 10^{6}$ nm/s)  a linear fit works pretty well. Using  a linear fit $%
y=109.6+9.59\ln (x)$ for $F_{max1}$ and Eq. \ref{f_logV_eq} we obtain the
distance between N and TS1, $x_{u(NS \rightarrow TS1)}=3.76$ \AA $\,$ which is comparable to
the experimental values  of 2.5--3.0 \AA \cite%
{Carrion-Vasquez_PNAS99,Rief_Science97}.  A nonlinear fit works for  a much
wider interval (Fig. \ref{unfold_barrier_CV}). Using Eq.\ref{Dudko_eq} with $\nu=1/2$  we
 obtain $x_{u(NS \rightarrow TS1)}=6.68\mathring{A}$ and $\Delta G_{1}^{\ddagger }=32.48k_{B}T
$ for $F_{max1}$. Similar results for $x_{u(NS \rightarrow TS1)}$ and  $\Delta G_{1}^{\ddagger }$  
 were found using  a fit 
with $\nu=2/3$ which works for  a slightly narrower interval (results not shown).
The distance to  the ransition state, $x_{u(NS \rightarrow TS1)}$,  
 based on the nonlinear theory 
is close to  the experimental value $x_{u}=5.9\mathring{A}$ reported by Williams 
\emph{et al} \cite{Williams_Nature03}.
Extrapolating results to the pulling speed $v=200$
nm/s used in the experiments,  we get  $F_{max1}\approx 160$ pN (Fig. \ref%
{unfold_barrier_CV}), which agrees  quite well with  the $F_{\text{max}}\approx
200$ pN obtained by stretching  a polyprotein of identical I27 domains in  the AFM experiment \cite{Marszalek_Nature99}.

In the case of $F_{max2}$, the linear fit gives $x_{u(IS \rightarrow TS2)}=2.95$ \AA , while
from the nonlinear theory we obtain $x_{u(IS \rightarrow TS2)}=3.88\mathring{A}$ and $%
\Delta G_{2}^{\ddagger }=9.22k_{B}T$. This result is consistent with the
previously reported values obtained by constant force Go-model simulations 
\cite{MSLi_JCP08}.

As stated above, the number of trajectories without the second peak
increases as pulling speed decrease s. Using the results presented in
Fig. \ref{unfold_barrier_CV} we can roughly estimate  the $v$ below which the
second maximum disappears. Using $y=-90.332+12.108\ln (x)$, we estimated the
second peak to be zero at  a pulling speed  of $v=1738$ nm/s. It is worth
noting that  the extrapolated value of $F_{max2}$ at the upper limit of AFM
pulling speed, $v=10^{4}$ nm/s , is only $21$ pN which cannot be
detected against the fluctuation background which could be as high as 30 pN
(Fig. 3 in Ref. \cite{Rief_Science97}). Thus, it is clear why the second
peak is not observed under experimental conditions. In addition,  the Go model
results suggest that in  the high pulling speed regime, $v>\sim 10^{4}$ nm/s ,
unfolding becomes three-state.  The applied force not only lower s the
energy barrier, but also leads to an additional transition state
not  found at low pulling speeds.

\subsection*{Switch in unfolding pathways and applicability of Bell's theory}

 It is noted that, the pulling speed $v\approx 10^{6}$ nm/s, at which we observed
the switch in  the unfolding pathway, falls within the interval $\sim
10^{6}-10^{7}$ nm/s. On the other hand, above this loading rate the linear
fitting based on Bell's theory ceases to work. Thus, the switch in  the 
unfolding pathway might be interpreted as the sign that Bell's theory is
 applicable  no more. This phenomenon was also observed for protein DDFLN4 
\cite{Li_JCP09} but the reason behind it was not discussed. Here we address
this problem in more detail. At weak forces ( low pulling speeds) the
protein experiences the action of the external force uniformly along the
chain and the secondary structure that has the weakest interaction with the
rest would unfold first and so all. However, for the finite
propagation speed of a perturbation caused by the force, the situation
 changes if a strong force is applied. In this case, it is not
necessary that the weakest part unfolds first if it is located far from the
point where  the force is applied. The secondary structure at the termin us
pulled by  the force may be detached earlier \cite{Li_JCP09}. Thus, the switch in 
mechanically unfolding pathways is associated with the
crossover from the weak force to the strong force  regime. On
the other hand, Bell's theory, which is valid at low forces, is
expected to fail in the strong force regime. Therefore, to  the best  of our knowledge  for the first time, we observe 
  a relationship between
the switch in  the unfolding pathway in Go models and the range of applicability
of Bell's theory. This observation is general and should work for other
models. It would be interesting to demonstrate this for other proteins using
Go and more precise modeling. 

It should be stressed that in our simulations
the switch in unfolding pathways occurs at pulling speeds of 10$^6$-10$^7$ nm/s.
In this regime the deviation from Bell behavior may be captured by the 1D theory \cite{Schlierf_BJ06,Dudko_PRL06}.
Combing  the experimental data with results obtained from high speed pulling simulations Schulten \emph{et al} 
 suggested that Bell's theory is violated at 
$v \sim 10^8$ nm/s \cite{Gao_JMRCM02} but they did not explicitly
show the change in pathways.
There exists another lower-limit of applicability of Bell's theory at very small
forces close to 0 \cite{Schlierf_BJ10, Yew_PRE10}. Here the change in
unfolding pathways is observed at $v \sim 20$ nm/s. The dependence of  the unfolding force 
on $v$ cannot be fitted by either Bell's or 1D theory.
 The coexistence of force-induced and zero-force (thermal or denaturant)
 unfolding pathways\cite{Yew_JPCB08} at small forces leads to non-zero unfolding forces which deviate significantly  
from what is predicted by these theories\cite{Schlierf_BJ10}.  To
correctly describe this phenomenon one has to go beyond
1D models using the
multidimensional energy landscape or taking into account alternative
 unfolding pathways\cite{Schlierf_BJ10, Yew_PRE10, Suzuki_PRL10, Suzuki_JCP11, Kappel_PRL12}.

\section{Conclusions}

The key result in this paper is that the mechanical unfolding pathway of  theI27
domain probed by  a structure-based C$\alpha $-model is not consistent with
experimental observations even at low pulling speeds. Similar ly to other $%
\beta $-strand proteins (DDFLN4 and FNIII)\cite{Li_JCP09,
Mitternacht_BJ09}, the unfolding pathway depends on pulling speed: at large $%
v$  the C-termin us unfolds first, while at speeds close to the
experimental ones  the N-termin us unfolds before  the C-termin us.

Comparing our simulations with the previous  reports, we find that  a similar
deviation from experimentally detected unfolding pathways was observed
previously not only in  the Go-model \cite{Cieplak_Proteins02, MSLi_JCP08} but
also in  the all-atom implicit solvent simulations \cite{Li_JCP03}. Both models
neglect  the IS probing the artificial unfolding pathway. The slowest pulling
speed, $v=2.5\times 10^{4}$nm/s, used in our Go-model simulation corresponds
to the upper limit of loading rates used in AFM experiments, $v=10^{4}$nm/s 
\cite{Carrion-Vasquez_PNAS99} and is nearly two orders of magnitude slower
than previously reported \cite{Cieplak_Proteins02}. It turns out that
probing the wrong pathway by  the Go-model is not related to the fast pulling,
but rather indicates more fundamental problems. The failure of  the Go-model to
detect the first transition state (or  the intermediate state), which
should be associated with a loss of native interactions comes as  a surprise.
This finding is valuable as the unfolding by  an external force is believed to
be solely governed by  the native topology of proteins. Thus, one has to maintain
a healthy skepticism about systematic studies based on Go-models\cite%
{Sulkowska_JPCM07, MSLi_BJ07a}. Although Go-models ha ve been proved to
provide reasonable estimates for mechanostability properties \cite%
{Brockwell_NSB03, Cieplak_JCP05, MSLi_BJ07, MSLi_PNAS06, Valbuena_PNAS09,
Arad_BJ10, Kumar_PR10}, there is no guarantee that they  can solve unfolding pathways. Benchmarking of simulation results on the
molecular basis of protein mechanostability by experiment is necessary to
make sure that properties measured in  an experiment are properly
reproduced by simulation.

The inclusion of more realistic interactions and explicit solvent 
influence into the model would help to  obtain proper pathways\cite%
{Lee_Structure09}. However, at present, deciphering  the unfolding FEL of long
proteins by all-atom simulations with explicit water is still
computationally prohibitive. Developing a model which is still computationally feasible and at  the same time takes into account 
 the contribution of side-chain and non-native interactions  would be of great help.
One of the possibilities 
is to use combination of
structure-based models with 
more realistic interactions
described by
statistics-based potentials\cite{Miyazawa_Macromolecules85, Kolinski_JCP93}.
We are  now using such a combination of  the Go-model with CABS software \cite%
{Kolinski_ABP04} to test the effectiveness of this idea in ongoing
simulations.

\clearpage

\begin{acknowledgments}
We thank M. Jamroz and A.M. Gabovich for helpful discussions. 
AK and MK would like to acknowledge support from the Foundation for Polish
Science TEAM project (TEAM/2011-7/6) cofinanced by the European Regional
Development Fund operated within the Innovative Economy Operational Program and from Polish Ministry of Science and 
Higher Education Grant No. IP2012 016872. MSL is supported by Narodowe Centrum Nauki in Poland (grant No
2011/01/B/NZ1/01622). CKH is supported by National Science Council  in
Taiwan under grant number NSC-100-2923-M-001-003-MY3 and National Center for
Theoretical Sciences in Taiwan. Allocation of CPU time at the supercomputer center TASK in Gdansk
(Poland) is highly appreciated.
\end{acknowledgments}

\clearpage



\clearpage
\noindent

{\huge Tables}

\begin{table}[!ht]
\begin{tabular}{|c|c|c|c|}
\hline
Pulling velocity (nm/s) & $F_{max1}$ & $F_{max2}$ & Number of Trajectories
\\ \hline
$v_1=2.5\times10^4$ & 210.5$\pm$7.3 & 36.6$\pm$8.5 & 50 \\ \hline
$v_2=5.76\times10^4$ & 215.4$\pm$13.3 & 44.2$\pm$12.3 & 50 \\ \hline
$v_3=1.29\times10^5$ & 221.0$\pm$18.3 & 50.7$\pm$11.4 & 50 \\ \hline
$v_4=2.88\times10^5$ & 224.3$\pm$24.1 & 57.8$\pm$16.2 & 50 \\ \hline
$v_5=6.44\times10^5$ & 232.4$\pm$27.7 & 64.1$\pm$17.1 & 50 \\ \hline
$v_6=1.44\times10^6$ & 253.7$\pm$29.3 & 80.7$\pm$23.5 & 50 \\ \hline
$v_7=3.22\times10^6$ & 264.4$\pm$29.7 & 98.8$\pm$29.5 & 50 \\ \hline
$v_8=7.2\times10^6$ & 288.1$\pm$33.2 & 119.9$\pm$31.6 & 50 \\ \hline
$v_9=1.6\times10^7$ & 352.1$\pm$48.3 & 178.9$\pm$44.2 & 50 \\ \hline
$v_{10}=3.6\times10^7$ & 422.6$\pm$57.2 & 283.2$\pm$34.2 & 50 \\ \hline
$v_{11}=8.05\times10^7$ & 549.8$\pm$52.4 & 448.0$\pm$52.3 & 50 \\ \hline
\end{tabular}
\linespread{0.8} \vspace{3 mm} 
\caption{List of pulling speeds used in simulations.
The upper limit for
pulling speed used in  the AFM experiment is $10^{4}$ nm/s
(taken from Ref.%
\protect\cite{Carrion-Vasquez_PNAS99})}

\label{sim_details1}
\end{table}

\clearpage

\noindent
{\huge Figure Captions}

\begin{figure}
\includegraphics[width=6.3in]{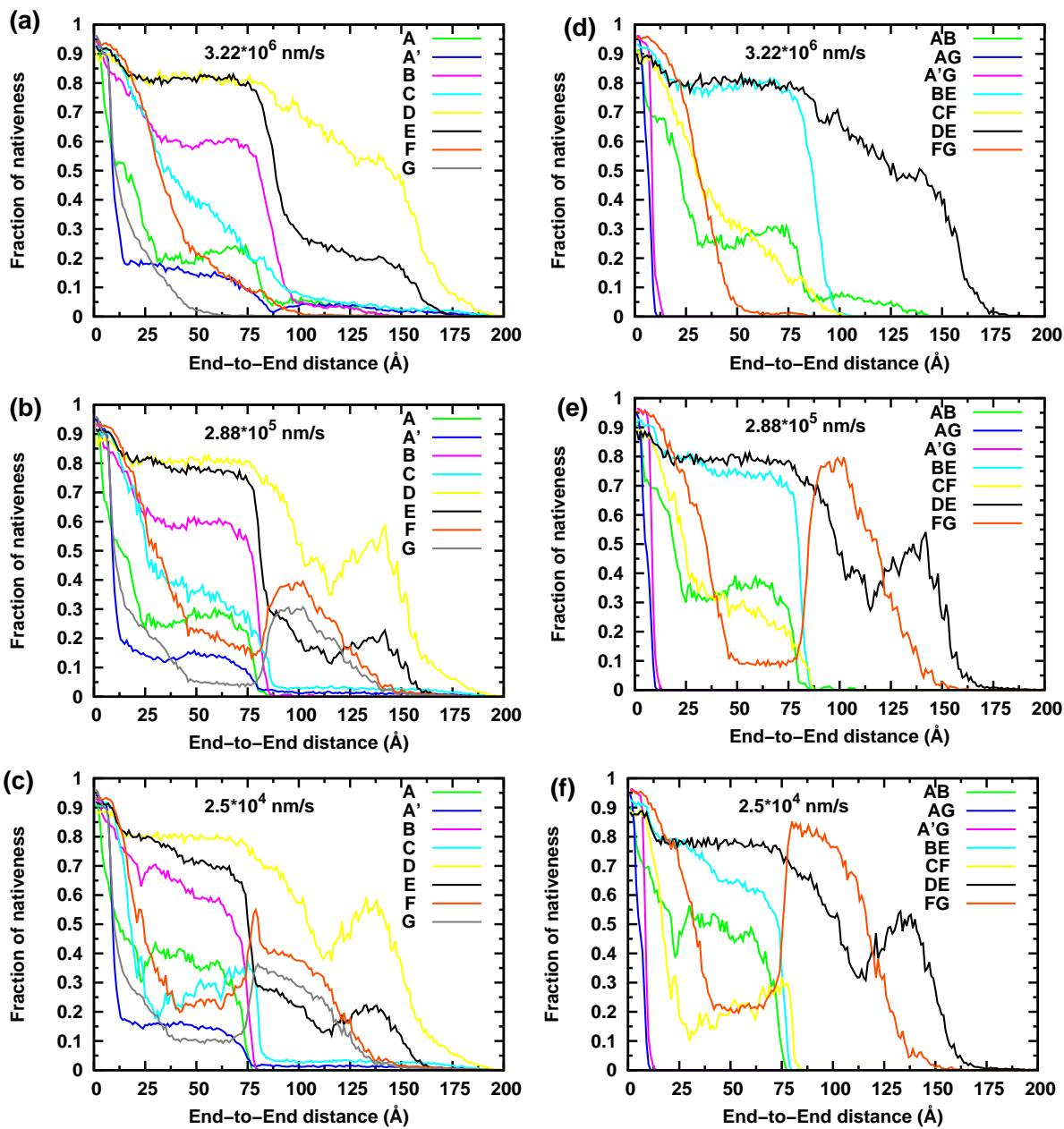} 
\caption{ End-to-end distance dependence of averaged fractions of native
		contacts. Native contacts are formed by 8 $\protect\beta $-strands marked  in Fig. \protect\ref%
		{Fig1_NS} at different loading rates. Clearly, the unfolding at high forces
		starts from  the C-termin us detaching G-strand first. In contrast, at low forces
		 the A and A' strands are unfolded first, but it should be noted that  the extension
		at which complete detachment of  the A strand takes place is rather large, $75~%
		\mathrm{\mathring{A}}$.}
\label{strands_pairs_Go}
\end{figure}

\begin{description}

\item[Fig.~1:]  Native state conformation of  the I27 domain of titin (PDB ID: 1TIT).
		There are 8 $\protect\beta$-strands: A (4-7), A' (11-15), B (18-25), C
		(32-36), D (47-52), E (55-61), F(69-75) and G (78-88).  (a) PDB structure in
		 cartoon representation. (b) Schematic view of  the same structure. Red and black $%
		\protect\beta$-strands belong to different $\protect\beta$-sheets. N- and C-terminal residues are marked N and C, respectively.  
\item[Fig.~2:]  The $f-T$ phase diagram obtained by the extended histogram method.
		The results were averaged over 30 trajectories. The vertical dashed line
		marks $T=0.42\protect\epsilon _{H}/k_{B}=280$ K at which  most of our
		calculations have been performed.
\item[Fig.~3:]   Averaged force-extension profiles of  the I27 titin domain. The results obtained by Go-model simulations
		performed at different pulling speeds  are shown  on the right. The inset shows an
		enlargement of the starting  region. Results have been averaged over 50
		trajectories. Values of pulling rates are given in Table \protect\ref%
		{sim_details1}.
\item[Fig.~4:]  Individual force-extension profiles of  the I27 titin domain. Four individual trajectories at $%
		2.5\times10^4$ nm/s obtained by Go-model simulations are presented. There is no second
		peak in two trajectories.
\item[Fig.~5:]  End-to-end distance dependence of averaged fractions of native
		contacts. Native contacts are formed by 8 $\protect\beta $-strands marked  in Fig. \protect\ref%
		{Fig1_NS} at different loading rates. Clearly, the unfolding at high forces
		starts from  the C-termin us detaching G-strand first. In contrast, at low forces
		 the A and A' strands are unfolded first, but it should be noted that  the extension
		at which complete detachment of  the A strand takes place is rather large, $75~%
		\mathrm{\mathring{A}}$.
\item[Fig.~6:]  End-to-end distance dependence of averaged fractions of native
		contacts. Same as Fig. \ref{strands_pairs_Go} but for the end-to-end distance up to 15 $\AA$.
		Within $15\mathring{A}$  the detachment of  the A strand out of  the protein core is not observed for
		any  speed studied.
\item[Fig.~7:] Typical unfolding pathway of titin from Go-model simulation.  Green and blue squares mark AB and A'G regions, respectively. The N-terminal
		residue is shown in magenta. (a) NS conformation. (b) Conformation at $4%
		\mathring{A}$ extension, contacts between AG are missing, both A'G and AB
		remained formed. (c) Conformation at $9\mathring{A}$ extension (after  the main
		peak), contacts between G with A and A' are broken, those between  the A and B
		strands are preserved. It should be emphasized that 100$\%$ trajectories at
		the beginning of unfolding (within $10\mathring{A}$) proceed via  the same
		pathway presented here regardless of the applied pulling rate. 
\item[Fig.~8:] Typical unfolding pathways of titin. (a-d) High pulling speed and (e-h) low pulling
		speed regions. The N-terminal residue is shown in magenta.
\item[Fig.~9:]  Force dependence on pulling speeds at $T=285K$. Force at  a given
		value of pulling speed is computed as  an average of maximum forces over 50
		trajectories. Grey boundaries of the polygon illustrate the interval of
		pulling rates used in  the AFM experiment. Black circles correspond to data for $%
		F_{max1}$. Solid and dashed black curves represent linear and nonlinear fits
		for $F_{max1}$. Similarly, red color is used for $F_{max2}$. Straight
		lines are fits to the Bell-Evans-Ritchie equation (Eq.\protect\ref{f_logV_eq}%
		), $y=109.6+9.59\ln (x)$ and $y=-90.332+12.108\ln (x)$ for $F_{max1}$ and $%
		F_{max2}$, respectively. Using  a linear fit we  found $x_{u(N\rightarrow
		TS1)}=3.76\mathring{A}$ and $x_{u(IS\rightarrow DS)}=2.95\mathring{A}$ for $%
		F_{max1}$ and $F_{max2}$, respectively. From  a nonlinear fit and Eq.\protect
		\ref{Dudko_eq} we got $x_{u}=6.68\mathring{A}$ and $\Delta G^{\ddagger
		}=32.48k_{B}T$ for $F_{max1}$ and $x_{u}=3.88\mathring{A}$ and $\Delta
		G^{\ddagger }=9.22k_{B}T$ for $F_{max2}$. Extrapolation to  the experimental
		pulling speed, $v=200$ nm/s, gives  a negative value of $F_{max2}$ regardless
		of the fit  used. Extrapolated values of $F_{max1}$ to $200$ nm/s are 160 and
		152 pN using linear and nonlinear fits, respectively.
		
\end{description}

\clearpage


\clearpage


\clearpage


\clearpage


\clearpage


\clearpage


\begin{figure}
\includegraphics[width=6.3in]{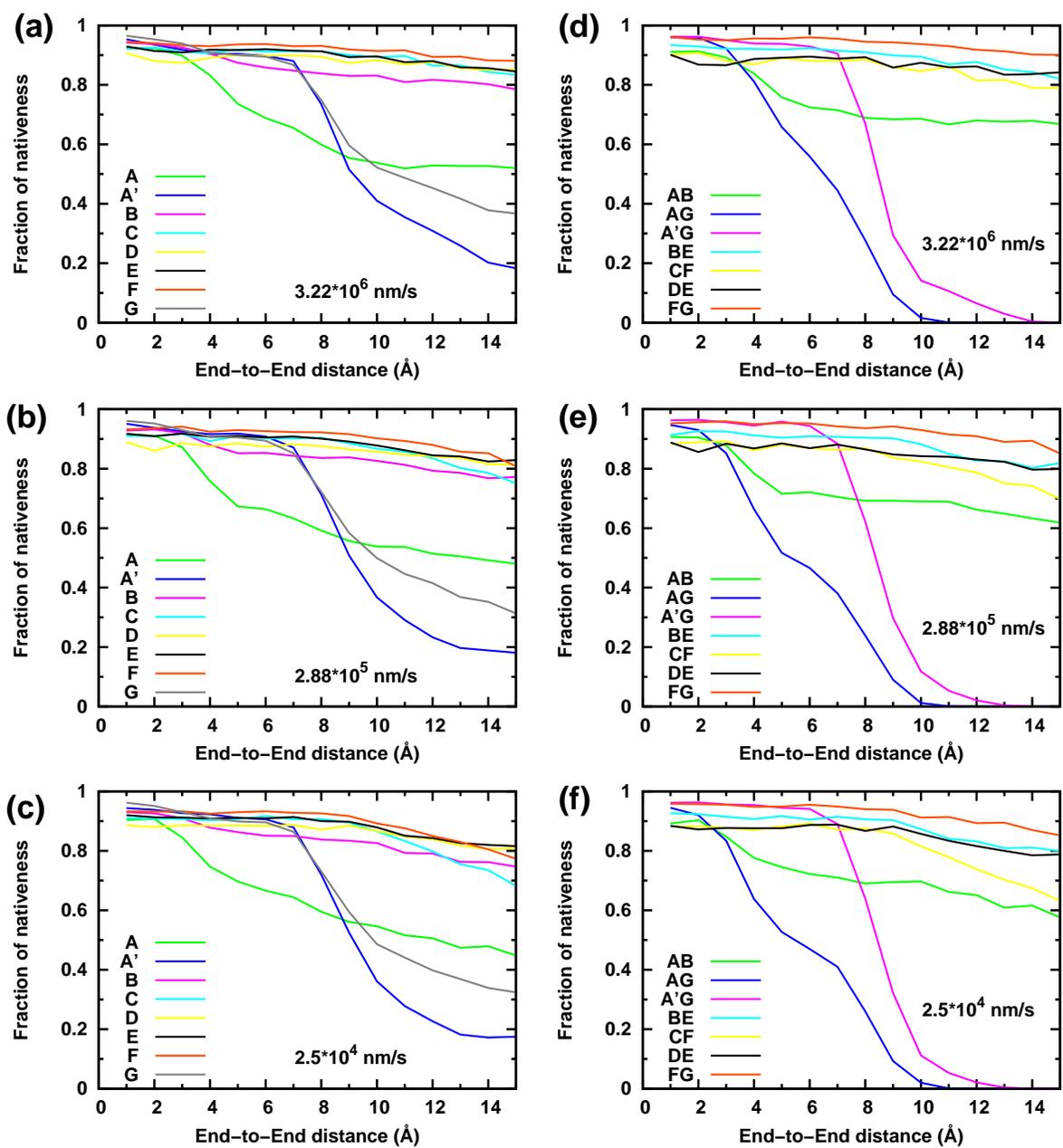} 
\caption{ End-to-end distance dependence of averaged fractions of native
		contacts. Same as Fig. \ref{strands_pairs_Go} but for the end-to-end distance up to 15 $\AA$.
		Within $15\mathring{A}$  the detachment of  the A strand out of  the protein core is not observed for
		any  speed studied.}
\label{inset_pairs_strands_Go}
\end{figure}

\clearpage


\begin{figure}
\includegraphics[width=6.3in]{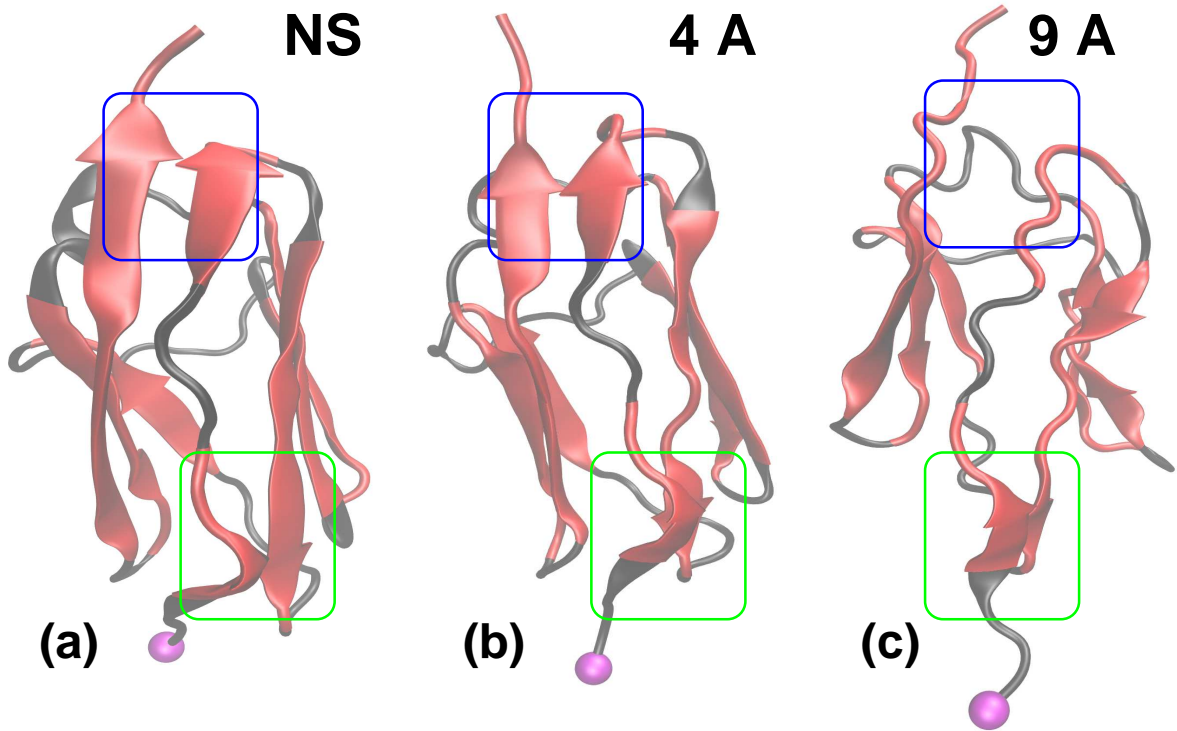} 
\caption{Typical unfolding pathway of titin from Go-model simulation. 
 Green and blue squares mark AB and A'G regions, respectively. The N-terminal
residue is shown in magenta. (a) NS conformation. (b) Conformation at $4%
\mathring{A}$ extension, contacts between AG are missing, both A'G and AB
remained formed. (c) Conformation at $9\mathring{A}$ extension (after  the main
peak), contacts between G with A and A' are broken, those between  the A and B
strands are preserved. It should be emphasized that 100$\%$ trajectories at
the beginning of unfolding (within $10\mathring{A}$) proceed via  the same
pathway presented here regardless of the applied pulling rate. }
\label{Go_model_path}
\end{figure}

\clearpage


\begin{figure}
\includegraphics[width=5.2in]{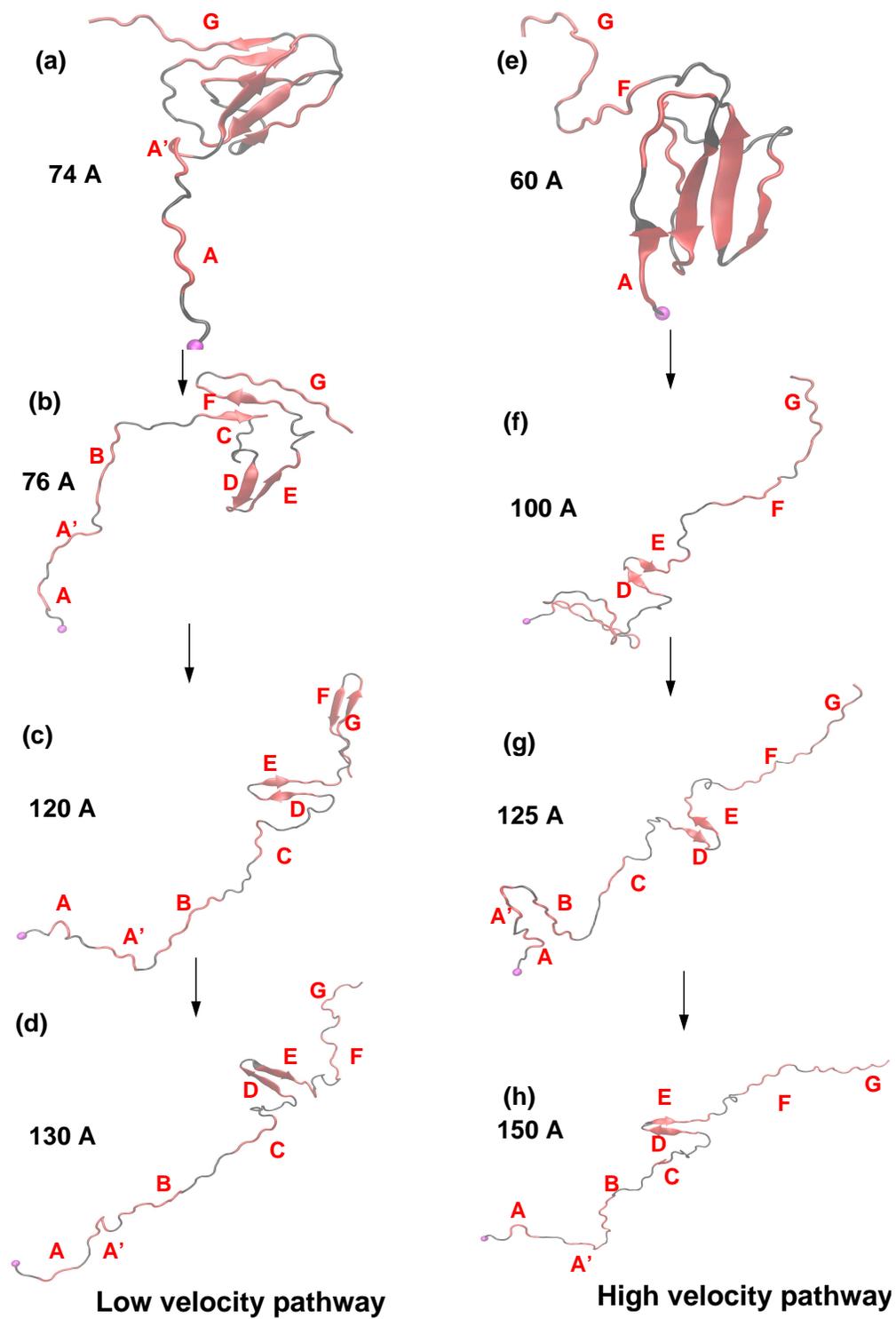} 
\caption{Typical unfolding pathways of titin. (a-d) High pulling speed and (e-h) low pulling
speed regions. The N-terminal residue is shown in magenta.}
\label{Go_model_path_all}
\end{figure}

\clearpage


\begin{figure}
\includegraphics[width=6.3in]{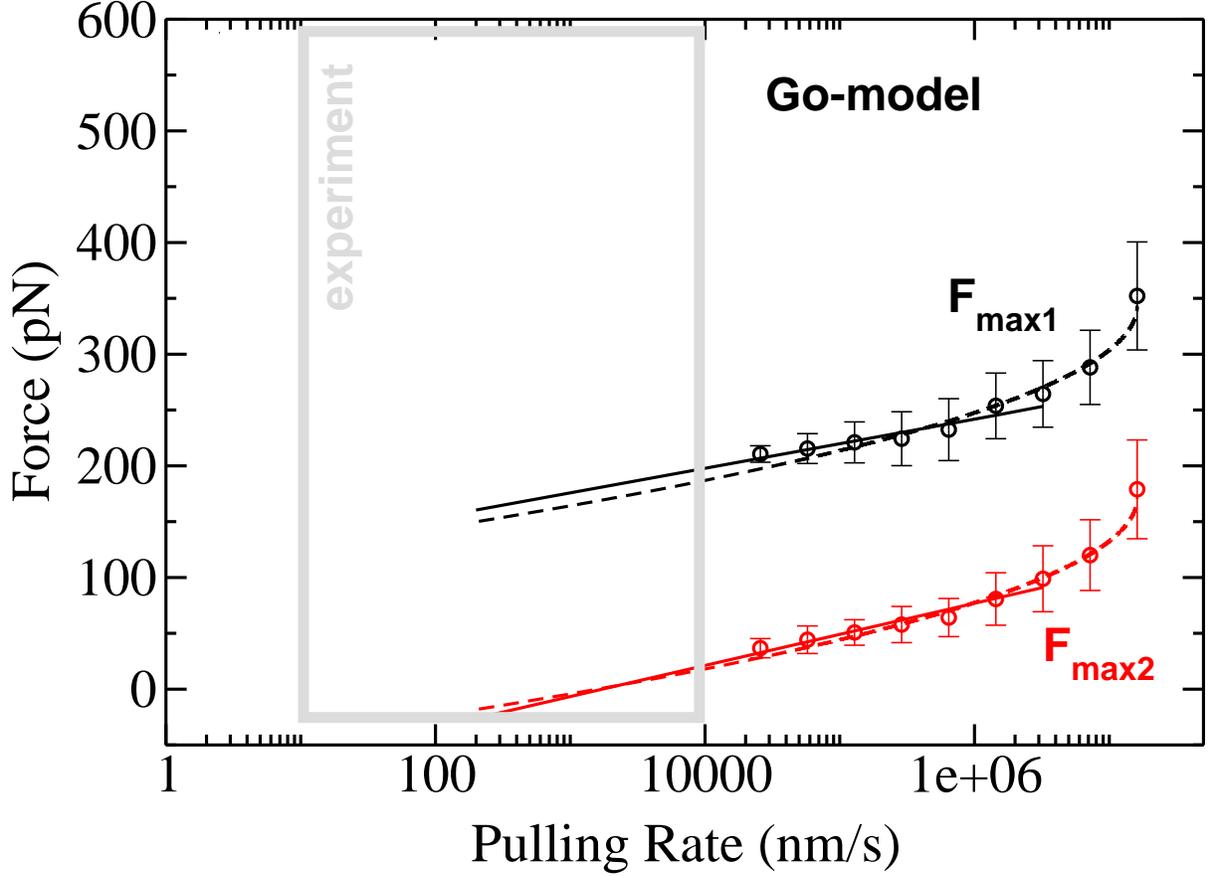} 
\caption{ Force dependence on pulling speeds at $T=285K$. Force at  a given
		value of pulling speed is computed as  an average of maximum forces over 50
		trajectories. Grey boundaries of the polygon illustrate the interval of
		pulling rates used in  the AFM experiment. Black circles correspond to data for $%
		F_{max1}$. Solid and dashed black curves represent linear and nonlinear fits
		for $F_{max1}$. Similarly, red color is used for $F_{max2}$. Straight
		lines are fits to the Bell-Evans-Ritchie equation (Eq.\protect\ref{f_logV_eq}%
		), $y=109.6+9.59\ln (x)$ and $y=-90.332+12.108\ln (x)$ for $F_{max1}$ and $%
		F_{max2}$, respectively. Using  a linear fit we  found $x_{u(N\rightarrow
		TS1)}=3.76\mathring{A}$ and $x_{u(IS\rightarrow DS)}=2.95\mathring{A}$ for $%
		F_{max1}$ and $F_{max2}$, respectively. From  a nonlinear fit and Eq.\protect
		\ref{Dudko_eq} we got $x_{u}=6.68\mathring{A}$ and $\Delta G^{\ddagger
		}=32.48k_{B}T$ for $F_{max1}$ and $x_{u}=3.88\mathring{A}$ and $\Delta
		G^{\ddagger }=9.22k_{B}T$ for $F_{max2}$. Extrapolation to  the experimental
		pulling speed, $v=200$ nm/s, gives  a negative value of $F_{max2}$ regardless
		of the fit  used. Extrapolated values of $F_{max1}$ to $200$ nm/s are 160 and
		152 pN using linear and nonlinear fits, respectively.}
\label{unfold_barrier_CV}
\end{figure}

\clearpage

\end{document}